# Probing Conformal Invariant of Non-unitary Two-Dimensional Systems by Central Spin Decoherence

Bo-Bo Wei

*School of Physics and Energy, Shenzhen University, Shenzhen, 518060, China*

**Universality classes of non-unitary critical theories in two-dimensions are characterized by the central charge. However, experimental determination of the central charge of a non-unitary critical theory has not been done before because of the intrinsic difficulty that complex parameters usually occur in non-unitary theory, which is not physical. Here we propose to extract the effective central charge of the non-unitary critical point of a two-dimensional lattice model from the quantum coherence measurement of a probe spin which is coupled to the lattice model. A recent discovery shows that quantum coherence of a probe spin which is coupled to a bath is proportional to the partition function of the bath with a complex parameter. Thus the effective central charge of a non-unitary conformal field theory may be extracted from quantum coherence measurement of a probe spin which is coupled to a bath. We have applied the method to the Yang-Lee edge singularity of the two-dimensional Ising model and extracted the effective central charge of the Yang-Lee edge singularity with good precision and tested other predictions of non-unitary conformal field theory. This work paves the way for the first experimental observation of the effective central charge of non-unitary conformal field theory.**

# Introduction

Scale invariance is one of the most intriguing features of statistical mechanics[1,2]. Local scale invariance (conformal invariance) at a critical point has been demonstrated to be remarkably powerful in two dimensions (2D)[3,4]. Universality classes of critical phenomena in two-dimensions are characterized by a single dimensionless number $c$, termed the central charge of the Virasoro algebra[5]. It was demonstrated that[6] unitarity constrains the values of $c<1$ to be quantized, $c = 1 - 6/[m(m+1)]$ with $m = 3, 4, 5, \cdots$. For such theories, the conformal dimensions of the primary fields are given by the Kac formula [7].

Non-unitary conformal field theory is often characterized by a negative effective central charge [8-10]. However, how to experimentally extract the effective central charge of the non-unitary critical point has not been known before because of the intrinsic difficulty that complex parameter occurs in non-unitary theory, which is not physical. It was found that the free energy at a conformal invariant critical point is linearly related to the effective central charge of the corresponding conformal field theory [11,12,8]. Thus experimental measurement of the free energy with a complex parameter is essential for extracting the effective central charge of a non-unitary conformal field theory. A recent discovery shows that the partition function with a complex parameter is related to the central spin decoherence[13,14]. This implies that the effective central charge of a non-unitary theory may be extracted by measuring the quantum coherence of a probe spin coupled to the 2D critical theory.

In this article, we propose to extract the effective central charge of a non-unitary critical point from the quantum coherence measurement of a probe spin which is coupled to the critical model. We show that the effective central charge of the non-unitary conformal field theory is related to the quantum coherence of the probe spin. We have applied the method to the Yang-Lee edge singularity of the two-dimensional Ising model, and showed that the central charge of the Yang-Lee edge singularity could be extracted with good precision from central spin coherence measurement. Furthermore, we show that other predictions of non-unitary conformal field theory could be tested by measuring the central spin coherence.

# Results

**Fundamentals of Conformal Minimal Models**

The simplest conformal field theory are those having finite number of independent fields, so called minimal conformal field theory[4]. The central charges of the conformal minimal models satisfy [3,6]

$$c = 1 - \frac{6(p-p')^2}{pp'}, \qquad (1)$$

where $p$ and $p'$ are coprime integers. For $|p-p'|>1$, there always exist two integers $r_0, s_0$ satisfying that $r_0 p - s_0 p' = 1$, and the corresponding theory has a negative conformal dimension[6],

$$h_{r_0,s_0} = -\frac{(p-p')^2 - 1}{4pp'}. \qquad (2)$$

In such case, the associated conformal minimal model is non-unitary[8-10]. We shall briefly review some important predictions of the non-unitary conformal field theory.

**Free Energy at Non-unitary Conformal Field Theory in Two-Dimensions.**

Let us consider a parallelogram with vertices at $(0, 2\pi, 2\pi\tau, 2\pi(1+\tau))$. The complex number $\tau$ with $\text{Im}\,\tau > 0$ is the modular parameter of the parallelogram. A torus with periodic boundary condition can be constructed from a finite cylinder of length $2\pi\,\text{Im}\,\tau$ by joining the ends and performing a twist around the axis by $2\pi\,\text{Re}\,\tau$. Thus we have the partition function on the torus[2,4]

$$Z(\tau,\bar{\tau}) = \text{Tr}[q^{L_0 - c/24} \bar{q}^{\bar{L}_0 - c/24}]. \qquad (3)$$

Here $q = e^{2\pi i \tau}, \bar{q} = e^{-2\pi i/\tau}$ and $(L_0, \bar{L}_0)$ are the generators of Virasoro algebras. Modular invariance gives the behaviour of the partition function as $q \to 0$ [11,12]

$$Z = (q\bar{q})^{-c_{\text{eff}}/24}. \qquad (4)$$

Here $c_{\text{eff}} = c$ for the unitary case and $c_{\text{eff}} = c - 24 h_{r_0 s_0} = 1 - \frac{6}{pp'}$ for non-unitary theory.

Thus one has the free energy at critical point in an $L \times L'$ rectangular lattice with periodic boundary condition [8],

$$F = \ln Z_{LL'} = F_0 + \frac{\pi c_{\text{eff}}}{6}\left(x + \frac{1}{x}\right). \qquad (5)$$

Here $x \equiv L'/L$. For systems with $c_{\text{eff}} > 0$, if the area of a two dimensional lattice $S = L' \times L$ is fixed and the shape varies, $F$ is in Equation (5) is maximal for a square lattice $L' = L$ ($x = 1$). Thus there is a thermodynamic driving force for elongation of the domain and this may be understood due to an attraction of the walls of the rectangle with force per unit length inversely proportional to the square of their separation. From this perspective, the tendency to elongation should be a general geometric effect. From Equation (5), the free energy per particle is,

$$f = \lim_{L' \to \infty} \frac{\ln Z_{LL'}}{LL'} = f_0 + \frac{\pi c_{\text{eff}}}{6L^2}. \qquad (6)$$

Here the first term depends on the area and is non-universal. However the second term depends on the effective central charge $c_{\text{eff}}$, which is universal for conformal field theory. Equation (6) provides a method to the determination of central charge $c_{\text{eff}}$ using finite-size scaling methods of the free energy at conformal invariant critical point. However, non-unitary conformal field theory is usually related to a complex parameter. This implies that we have to measure the free energy with a complex parameter in order to test the predictions of non-unitary conformal field theory. Recently, the author and his collaborators found that the central spin coherence and the partition function with a complex parameter are deeply related [13,14]. For completeness, we shall review briefly the method in the next section.

**Partition Function with a Complex Parameter and Central Spin Decoherence**
To extract the central charge of non-unitary conformal invariant critical point, we should design a method for measuring the free energy with a complex parameter. Previous investigations show that the central spin coherence and partition function are deeply related[13-18], which we briefly explain below. Let us consider a general many-body system with Hamiltonian,

$$H(\lambda) = H_0 + \lambda H_1, \tag{7}$$

where $H_0$ and $H_1$ are two competing Hamiltonians and $\lambda$ is a control parameter of the system. We introduce a probe spin-1/2 (all termed central spin-1/2) coupled to the many-body system (bath), with probe bath interaction $H_I = -\eta S_z \otimes H_1$ and $\eta$ being a coupling constant between the probe spin and bath and $S_z = |\uparrow\rangle\langle\uparrow| - |\downarrow\rangle\langle\downarrow|$ being the Pauli matrix of the probe spin along $z$ direction. If we initialize the probe spin in a superposition state as $|\uparrow\rangle + |\downarrow\rangle$ and the bath at thermal equilibrium state with inverse temperature $\beta = 1/T$ ($k_B = 1$) described by $\rho_0 = e^{-\beta H(\lambda)} / Z(\beta, \lambda)$ with $Z(\beta, \lambda) = \text{Tr}[e^{-\beta H(\lambda)}]$ being the partition function of the bath. Then the probe spin and bath evolve in time together under Hamiltonian $H + H_I$. At time $t$, the quantum coherence of the probe spin, defined by $\langle S_+(t)\rangle \equiv \langle S_x(t)\rangle + i\langle S_y(t)\rangle$, has an intriguing form as[13,14],

$$\langle S_+(t)\rangle = \frac{Z(\beta, \lambda + i\eta t/\beta)}{Z(\beta, \lambda)}. \tag{8}$$

The denominator in the above equation is nonzero for real temperature $\beta$ and real control parameter $\lambda$. The numerator resembles the form of a partition function but with a complex control parameter, $\lambda + i\eta t/\beta$. The evolution time $t$ plays the role of the imaginary part of the control parameter. Equation (8) establishes the relation between partition function with a complex parameter and the central spin coherence, which leads to the first experimental observation of Lee-Yang zeros[16,17].

In experiment, the central spin coherence can be measured from the following steps:

(1). Initialize the system in the equilibrium state described by the Gibbs density matrix $\rho_0 = e^{-\beta H(\lambda)} / Z(\beta, \lambda)$ and the central spin in the $|\downarrow\rangle$ state;

(2). Apply a $\pi/2$ pulse along the $y$ direction to the central spin, which then transforms the central spin in a coherent superposition state as $(|\uparrow\rangle + |\downarrow\rangle)/\sqrt{2}$;

(3). The central spin and the system evolve together in time for a time interval $t$ and the evolution is governed by the total Hamiltonian $H(\lambda)+H_I$;

(4). Apply a $\pi/2$ pulse to the central spin along the *y* direction;

(5). Measure the average value of the central spin along *z* and *y* directions respectively, namely, $\langle S_z(t)\rangle$ and $\langle S_y(t)\rangle$, which are the real and imaginary part of the quantum coherence of the central spin $\langle S_+(t)\rangle$.

Combining Equation (6) and Equation (8), the free energy at a non-unitary critical point can be written as

$$\begin{aligned}
f_{LL'}(t_c) &= \lim_{L'\to\infty}\frac{\ln\left(Z_{LL'}(\beta,\lambda+i\eta t_c/\beta)\right)}{LL'}, \\
&= \lim_{L'\to\infty}\frac{\ln\left(\langle S_+(t_c)\rangle Z_{LL'}(\beta,\lambda)\right)}{LL'}, \\
&= \lim_{L'\to\infty}\frac{\ln\left(\langle S_+(t_c)\rangle\right)}{LL'} + \lim_{L'\to\infty}\frac{\ln\left(Z_{LL'}(\beta,\lambda)\right)}{LL'}, \\
&= f_0 + \frac{\pi c_{\text{eff}}}{6L^2}.
\end{aligned} \tag{9}$$

Here $\lambda+i\eta t_c/\beta$ is the critical point of the non-unitary conformal field theory in two-dimensions. In Equation (9), $\lim_{L'\to\infty}\frac{\ln(Z_{LL'}(\beta,\lambda))}{LL'}$ is *free energy per particle* away from critical point and thus can be considered as a constant. Therefore we have

$$\lim_{L'\to\infty}\frac{\ln\left(\langle S_+(t_c)\rangle\right)}{LL'} = f_0' + \frac{\pi c_{\text{eff}}}{6L^2}. \tag{10}$$

Equation (10) is the central result of this work. It implies that the central spin coherence at the critical point of a non-unitary conformal field theory is related to the effective central charge of the non-unitary conformal field theory. Therefore, one may be able to experimentally extract the effective central charge of a non-unitary conformal field theory by measuring the quantum coherence of a probe spin which is coupled to a critical 2D lattice.

The effective central charge of a non-unitary conformal field theory can be extracted experimentally from the following steps. Firstly, one measures the quantum coherence of a probe spin coupled to a 2D lattice with different sizes as a function of time $\langle S_+(t)\rangle$. Secondly, doing finite-size scaling analysis [18], one can get the critical point of the conformal field theory, that is $\lambda + i\eta t_c/\beta$. This can be achieved by finite size scaling analysis [18]. Third, getting quantum coherence at the critical point $\langle S_+(t_c)\rangle$ for different lattice sizes and fitting these data points with a linear function of $1/L^2$, the coefficient of $1/L^2$ tells us the effective central charge of the non-unitary conformal field theory according to Equation (10). Because central spin coherence is directly experimentally measurable [19-23], we can extract the effective central charge of non-unitary critical point in two-dimension from central spin coherence measurement according to above procedures. To demonstrate the feasibility of the proposal, we shall study the Yang-Lee edge singularity [24,25] in two-dimensional Ising model, which is a typical example of non-unitary conformal invariant critical point as pointed out by Cardy [26].

Besides, numerical methods such as the thermodynamic Bethe ansatz approach[27-29] has been used successfully to compute the free energy of both unitary and non-unitary critical points. Recently, the effective central charge also plays a role is in the study of the entanglement entropy of non-unitary conformal field theories [30-32]. This provides a new physical quantity that can be evaluated numerically and from which the effective central charge may be extracted.

**Discussion.**
**Yang-Lee Edge Singularity in Two-dimensional Ising Model**
To illustrate the above idea, we study the Yang-Lee edge singularity in two-dimensional (2D) Ising model. The Hamiltonian of the square lattice Ising model with *N* columns and *M* rows is

$$H = -J\sum_{i=1}^{N}\sum_{j=1}^{M}(\sigma_{i,j}\sigma_{i+1,j} + \sigma_{i,j}\sigma_{i,j+1}) - h\sum_{i=1}^{N}\sum_{j=1}^{M}\sigma_{i,j}. \qquad (11)$$

Here $\sigma_{i,j} = \pm 1$ is the spin in the site (*i*,*j*) and *J* is the ferromagnetic coupling constant between nearest-neighbor spins in the lattice and *h* is the magnetic field experienced by all the spins and periodic boundary conditions are applied. We set *J*=1 and take it as the basic unit of the energy scale. The 2D Ising model at zero magnetic field *h*=0 has been exactly solved by Onsager in 1944 [33] and there is a finite temperature phase transitions at $\beta_c = \ln(1+\sqrt{5})/2$ [33,34]. For 2D Ising model under a finite magnetic field, there is no exact solution available but one can map the problem into 1D quantum Ising model with both longitudinal and transverse field by transfer matrix method [35,36].

At any temperature above the critical temperature, $\beta < \beta_c$, the 2D Ising model is critical for a purely imaginary magnetic field $h_c = \pm i h_c(\beta)$. This critical point with complex magnetic field is termed the Yang-Lee edge singularity[24,25]. As pointed out by Cardy [26], the Yang-Lee edge singularity at 2D Ising model is a non-unitary conformal invariant critical point and corresponds to the minimal conformal field theory with $p = 5, p' = 2, c = -22/5$. There are only two primary fields, the identity $I$ with $h_{11} = 0$ and the scalar field with $h_{12} = -1/5$. The only modular invariant partition function build out of these two fields is [8,9]

$$Z(q,\bar{q}) = (q\bar{q})^{11/60}\left(|\chi_{11}(q)|^2 + |\chi_{12}(q)|^2\right). \tag{12}$$

As $q \to 0$, we have,

$$\lim_{M\to\infty} \frac{\ln Z_{MN}}{MN} = f_0 + \frac{\pi}{6}\frac{c_{\text{eff}}}{N^2}. \tag{13}$$

Here $c_{\text{eff}} = c - 24h_{12} = 1 - 6/pp' = 2/5$ for the Yang-Lee edge singularity in 2D lattices. Since Yang-Lee edge singularity occurs only for purely imaginary magnetic field for ferromagnetic Ising model, we have from Equation (10)

$$\lim_{L'\to\infty} \frac{\ln\left(\langle S_+(t_c)\rangle\right)}{LL'} = f_0' + \frac{\pi c_{\text{eff}}}{6L^2}. \tag{14}$$

Here $\langle S_+(t_c)\rangle = Z_{MN}(\beta, ih_c(\beta))/Z_{MN}(\beta, 0)$ is the central spin decoherence at Yang-Lee edge singularity point and $t_c = \beta h_c(\beta)/\eta$.

First, we study the Yang-Lee edge singularity of the 2D rectangular lattice Ising model with $M \times N$ spins at inverse temperature $\beta = 0.2$. The Yang-Lee edge critical point for $\beta = 0.2$ located at $t_c = 0.257$, which is obtained from the finite size scaling method at Yang-Lee edge singularity[18]. We then show the central spin coherence at the Yang-Lee edge singularity in 2D Ising model at $\beta = 0.2$ as a function of $N$ for different lattice sizes when *M*=80 fixed in Figure 1(a). From Equation(14), we know that for a rectangular lattice with $M \times N$ spins, if $M \gg N$, the logarithm of the central spin coherence per spin at the Yang-Lee edge singularity is a linear function of $1/N^2$ with the slop being $\pi c_{eff}/6$. We then make a linear fit of the logarithm of the central spin coherence per bath spin at the Yang-Lee edge singularity as a function of $1/N^2$ in Figure 1(b) and found that the slope is $0.210 \pm 0.001$. From Equation(14), we know that the effective central charge $c_{eff} \approx 0.40 \pm 0.01$. Thus the estimated effective central charge agrees with the exact solution $c_{eff} = 2/5$ perfectly.

Since Yang-Lee edge singularity occurs for any temperature above the critical temperature of the bulk system $\beta < \beta_c$. To test the claim, we study the Yang-Lee edge singularity at a different temperature $\beta = 0.3$. From the finite size scaling analysis, we know that the Yang-Lee edge critical point for $\beta = 0.3$ located at $t_c = 0.085$ [18]. Figure 2 (a) presents the central spin coherence at the Yang-Lee edge singularity in 2D Ising model at $\beta = 0.3$ as a function of *N* for different lattice sizes when *M*=80 fixed. We then make a linear fit of the logarithm of the central spin coherence per spin at the Yang-Lee edge singularity as a function of $1/N^2$ in Figure 2 (b) and found that the slope is $0.205 \pm 0.001$. From Equation (14), we know that the effective central charge $c_{eff} \approx 0.39 \pm 0.01$. Thus the estimated effective central charge agrees with the exact solution $c_{eff} = 2/5$.

We further test the effect of elongation of the 2D lattices in non-unitary conformal field theory corresponding to the two-dimensional critical point. Because of the intimate link between the central spin coherence and the free energy, it is sufficient to test the central spin coherence due to the elongation effect. Figure 3 shows the logarithm of the central spin coherence per bath spin at the Yang-Lee edge singularity of 2D Ising model at $\beta = 0.2$ for 2D rectangular lattices with lattice sizes being $2 \times 50, 4 \times 25, 5 \times 20, 10 \times 10$ (with same area $M \times N = 100$) as a function of the logarithm of the ratio of two edges of the rectangular lattice. First, one can see that the logarithm of the central spin coherence per bath spin is invariant under the modular transformation, interchanging of *M* and *N*, i.e. $x \leftrightarrow 1/x$. Second, the logarithm of the central spin coherence per bath spin of a 2D domain with a fixed area is maximum for square $x=1$ and decreases as $x$ becomes larger or smaller. This means there is a thermodynamic driving force for elongation of the a rectangular domain at the non-unitary conformal field theory.

## Summary

In summary, we show that the effective central charge of the non-unitary conformal invariant critical point of a two-dimensional lattice model can be extracted from the quantum coherence measurement of a probe spin which is coupled to the two-dimensional lattice model. Furthermore, we show that the other predictions by the non-unitary conformal field theory can also be tested from measuring the quantum coherence of the probe spin. Thus measuring the quantum coherence of a single probe spin provides a practical approach to studying the non-unitary conformal invariant two-dimensional interacting many-body systems and pave the way for the first experimentally verification of the predictions of non-unitary conformal field theory.

## Methods

The 2D Ising model without magnetic field can be mapped to a 1D Ising model with a transverse field by the transfer matrix method [35]. For a 2D Ising model in finite magnetic field, it was mapped by transfer matrix method to a 1D Ising model with both longitudinal field and transverse field [36], which can be numerically diagonalized. Therefore the partition function and hence the probe spin coherence for the 2D Ising model in finite magnetic field can be obtained.

**Acknowledgements:** B.B.W. was supported by National Natural Science Foundation of China (Grants No. 11604220) and the Startup Funds of Shenzhen University (Grants No. 2016018).

**Author Contributions**: B.B. W conceived the idea, performed the research and wrote the manuscript.

**Competing financial interests** The authors declare no competing financial interests.

**Correspondence** and requests for materials should be addressed to B.B.W. (bbwei@szu.edu.cn).

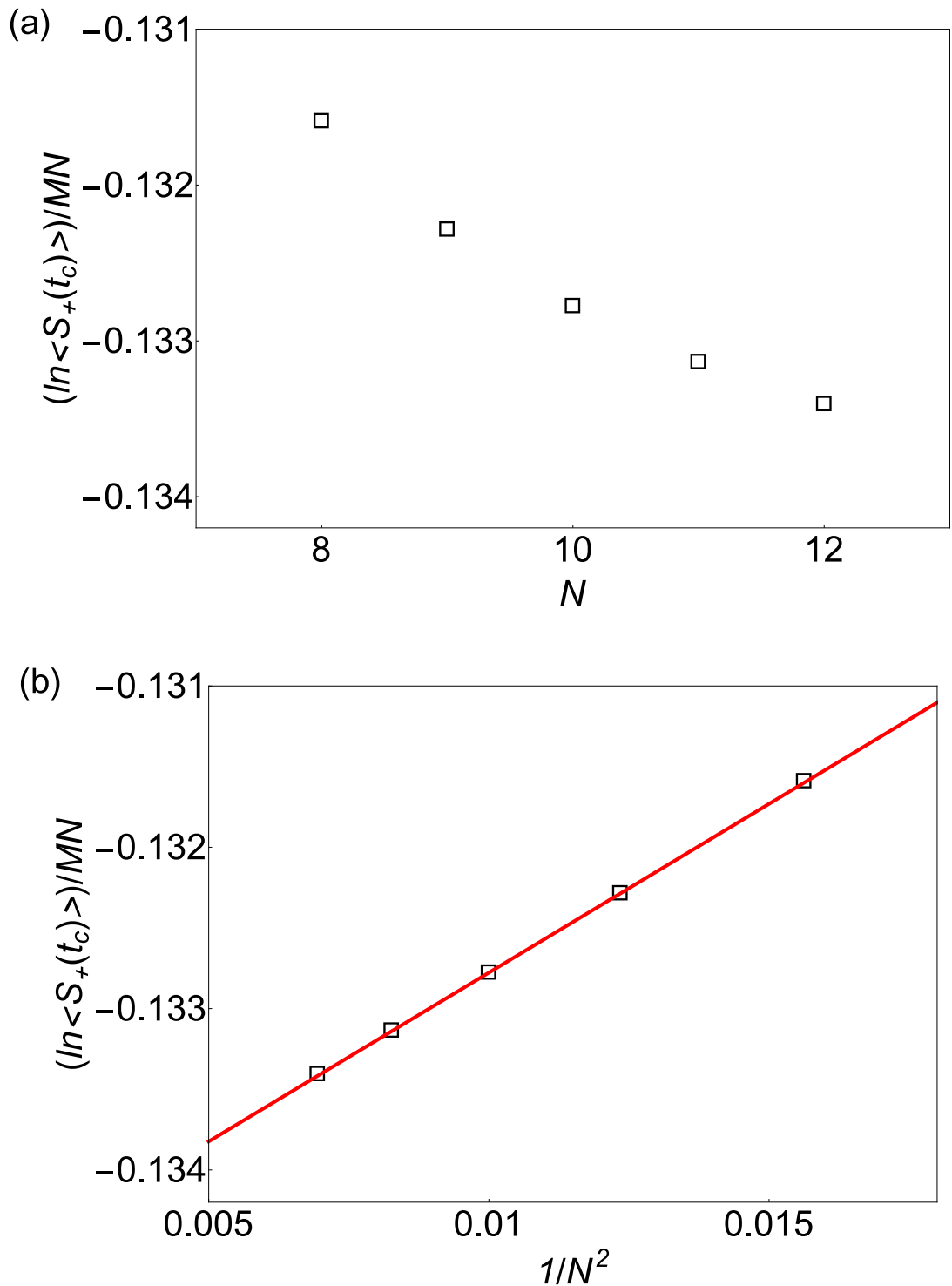

**Figure 1 | Extracting the effective central charge of the Yang-Lee edge singularity in 2D Ising model at inverse temperature** $\beta = 0.2$. (a). The logarithm of the central spin decoherence per bath spin (hollow square) at the Yang-Lee edge singularity in 2D Ising model of rectangular lattice with $M \times N$

spins as a function of the number of spin in one edge $N$, where the number of spins in the other edge of the rectangular lattice $M$, is fixed to be 80. (b). The logarithm of the central spin coherence per bath spin at the Yang-Lee edge singularity in 2D Ising model of rectangular lattice as a function of $1/N^2$ (hollow square). The red solid line is a linear fit of these data as a function of $1/N^2$. The slope of the solid line is 0.210 and thus the effective central charge is $0.210 \times 6/\pi \approx 0.40$.

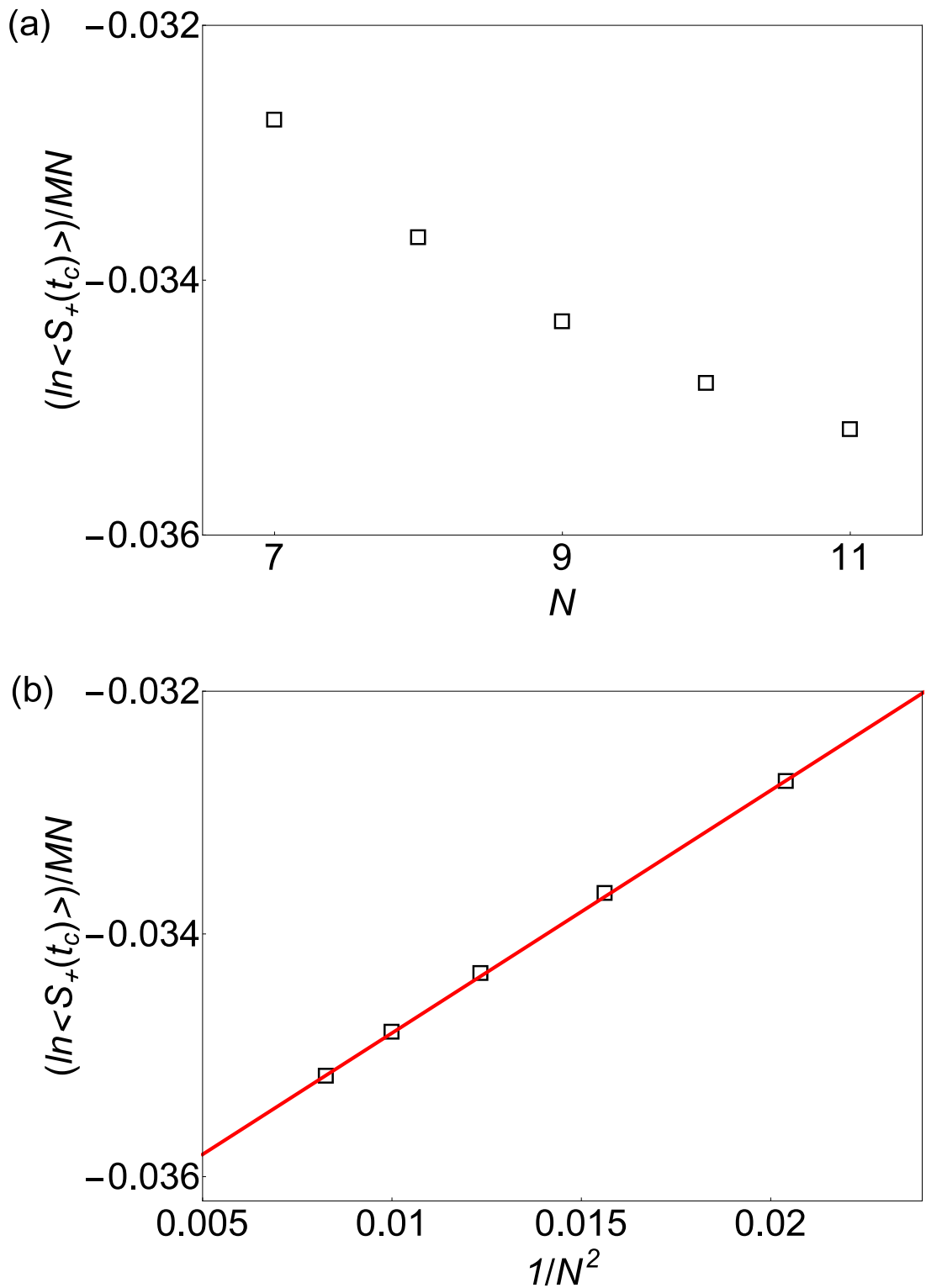

**Figure 2 | Extracting the effective central charge of the Yang-Lee edge singularity in 2D Ising model at inverse temperature** $\beta = 0.3$. (a). The logarithm of the central spin coherence per spin (hollow square) at the Yang-Lee edge singularity in 2D Ising model of rectangular lattice with $M \times N$ spins as a

function of the number of spin in one edge $N$, where the number of spins in the other edge of the rectangular lattice $M$, is fixed to be 80. (b). The logarithm of the central spin coherence per bath spin at the Yang-Lee edge singularity in 2D Ising model of rectangular lattice as a function of $1/N^2$ (hollow square). The red solid line is a linear fit of these data as a function of $1/N^2$. The slope of the solid line is 0.205 and thus the effective central charge is $0.205 \times 6/\pi \approx 0.39$.

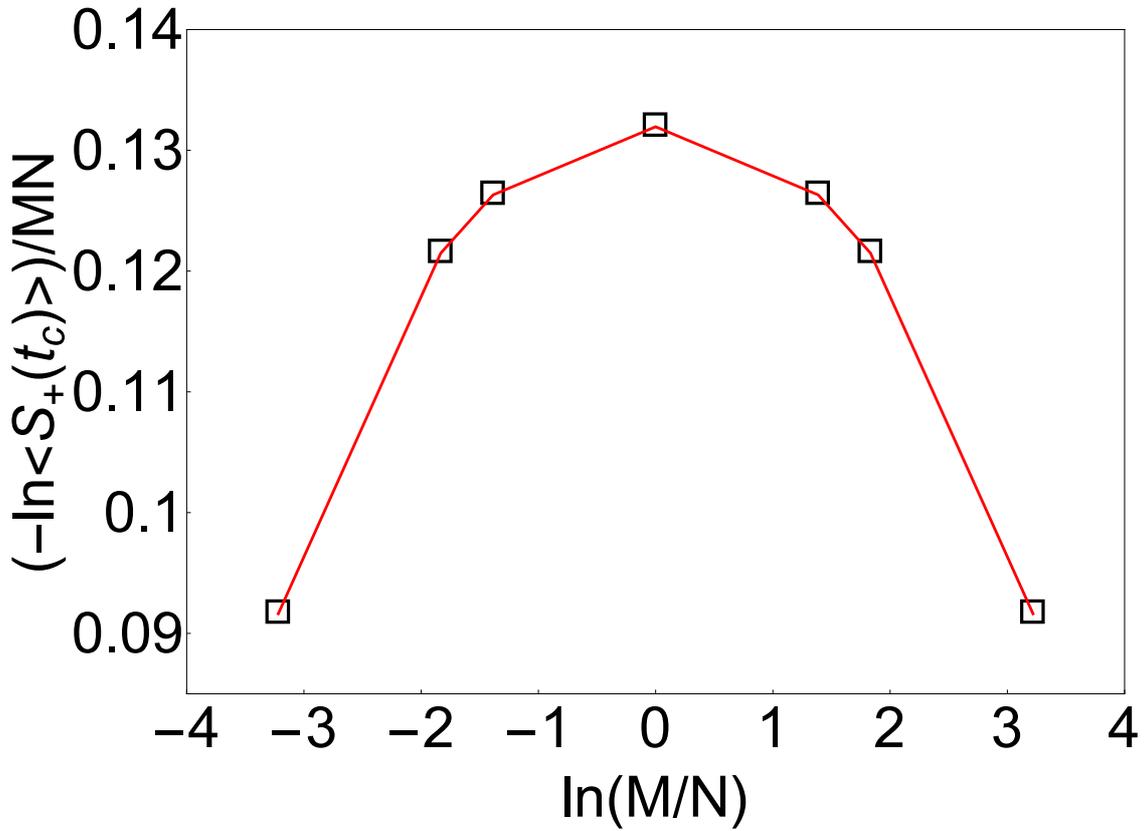

**Figure 3 | The effect of elongation of the 2D rectangular lattice at the Yang-Lee edge singularity in two-dimensional Ising model**. The vertical axis shows the logarithm of the central spin coherence per spin (hollow square) at the Yang-Lee edge singularity in 2D Ising model of rectangular lattices with the same area $M \times N = 100$ spins at inverse temperature $\beta = 0.2$ and the horizontal axis presents the logarithm of the ratio of the two edges of the rectangular lattice, $x = M/N$. All the data points presented have the same area $S = M \times N$.